\documentclass[twocolumn,aps,prd,amsmath,amssymb,preprintnumbers,longbibliography]{revtex4-1}
\usepackage{amsmath} \usepackage{amsfonts} \usepackage{amssymb}
\usepackage{bbm}
\usepackage{epsfig}
\usepackage{graphics}
\usepackage{graphicx}
\usepackage{titlesec}
\usepackage{mathtools}
\usepackage{environ}
\usepackage{dsfont}
\usepackage{todonotes}

\usepackage[colorlinks=true]{hyperref}
\hypersetup{urlcolor=black,linkcolor=black,citecolor=black}
\usepackage[toc,page]{appendix}

\makeatletter
\renewcommand*\env@matrix[1][c]{\hskip -\arraycolsep
  \let\@ifnextchar\new@ifnextchar
  \array{*\c@MaxMatrixCols #1}}
\makeatother

\newcommand{\be}{\begin{equation}}
\newcommand{\ee}{\end{equation}}
\newcommand{\ba}{\begin{eqnarray}}
\newcommand{\ea}{\end{eqnarray}}
\newcommand{\nn}{\nonumber}

\newcommand{\pt}{\partial_t}
\newcommand{\tr}{{\rm tr}}
\newcommand{\mn}{{\mu\nu}}
\newcommand{\rt}{{\rho\tau}}
\newcommand{\cl}{{\cal L}}

\titleformat{\subsection}[block]{\normalfont\bfseries}{\thesubsection.}{1ex}{}
\titlespacing{\subsection}{0pt}{10pt}{1pt}[0pt]
\titleformat*{\section}{\large\bfseries}
\renewcommand{\thesubsection}{\arabic{subsection}}

\newcommand{\p}{\partial}

\usepackage{braket}

\newcommand{\raw}{\rightarrow}

\begin{document}

\title[ ]{Infrared limit of quantum gravity}

\author{C. Wetterich}
\email{c.wetterich@thphys.uni-heidelberg.de}
\affiliation{Institut  f\"ur Theoretische Physik\\
Universit\"at Heidelberg\\
Philosophenweg 16, D-69120 Heidelberg}

\begin{abstract}

We explore the infrared limit of quantum gravity in presence of a cosmological constant or effective potential for scalar fields. For a positive effective scalar potential, one-loop perturbation theory around flat space is divergent due to an instability of the graviton propagator. Functional renormalization solves this problem by a flow of couplings avoiding instabilities. This leads to a graviton barrier limiting the maximal growth of the effective potential for large values of scalar fields. In the presence of this barrier, variable gravity with a field dependent Planck mass can solve the cosmological constant problem by a cosmological runaway solution. We discuss the naturalness of tiny values of the cosmological constant and cosmon mass due to a strong attraction towards an infrared fixed point. 

\end{abstract}

\maketitle

\section{Introduction}
\label{Introduction}

Functional renormalization for the effective average action or flowing action \cite{wetterich1993exact,BTW} aims for a description of quantum gravity from the shortest to the longest distances or wavelengths. An ultraviolet (UV) fixed point can define an asymptotically safe model \cite{weinberg1979ultraviolet} and render quantum gravity non-perturbatively renormalizable. We address here the infrared (IR) limit. The existence of a consistent IR-limit is a prerequisite for a consistent and reliable description of quantum gravity. Furthermore, a crossover between UV- and IR-fixed points in scalar-tensor theories of quantum gravity can provide a realistic cosmology with a single scalar field responsible both for early inflation and late dynamical dark energy \cite{wetterich2015inflation,rubio2017emergent}.

In the absence of a cosmological constant, perturbative quantum gravity can be treated as an effective low energy theory, expanded around flat space \cite{DONO1,DONO2}. A one loop computation yields rather mild non-local corrections to the inverse graviton propagator $\sim q^4\ln q^2$, to be compared with the classical contribution $\sim M^2q^2$, with $M$ the Planck mass. This type of computation can be extended to a negative cosmological constant. For a positive cosmological constant, however, the graviton propagator in flat space becomes tachyonic. A perturbative expansion around flat space is no longer possible, since the instability is associated with divergent momentum integrals.

Accordingly, already the pioneering work by Reuter \cite{reuter1998nonperturbative} on the functional renormalization flow of quantum gravity has revealed problems to achieve a proper IR-limit in case of a positive cosmological constant. The issue concerns an instability in the flow equation that needs to be controlled in a proper way \cite{RESAU2}. In turn, this instability is rooted in the perturbative instability of the flat-space graviton propagator in presence of a positive cosmological constant. The continued discussion of the role of the instability \cite{reuter2004quantum,reuter2009background,reuter2009conformal,nagy2013critical,christiansen2014fixed,christiansen2016global,christiansen2015local,
biemans2017quantum,denz2016towards,biemans2017renormalization} may have left the impression that results depend strongly on the method and truncation chosen and can therefore not be considered as reliable.

In ref.~\cite{CWGF} it was proposed that the use of a gauge invariant flow equation (for a single metric field) \cite{wetterich2016gaugeinvariant,CW17} leads to a conceptually simple solution of the infrared problem. The flow of couplings is such that the singularity is always avoided. In the infrared limit a positive cosmological constant or effective scalar potential is subject to strong non-perturbative renormalization effects. The flow entails a ``graviton barrier'' implying that for an increasing scalar field $\chi$ the effective potential $V(\chi)$ cannot increase faster than the squared $\chi$-dependent Planck mass $M^{2}(\chi)$. For a cosmology characterized by a runaway solution for which $\chi$ increases towards infinity in the infinite future, the observable dimensionless cosmological constant $V/M^{4}$ vanishes if $M^{2}$ diverges for $\chi\raw \infty$ \cite{wetterich1988cosmology}.

This constitutes a dynamical solution of the cosmological constant problem. In particular, for models of variable gravity \cite{wetterich2014variable}, with $M^{2}=\chi^{2}$ for $\chi\raw\infty$, the graviton barrier implies that the ratio $V/M^{4}$ has to vanish $\sim \chi^{-2}$ or faster. The presently still non-zero value of $V$ can constitute dynamical dark energy or quintessence \cite{wetterich1988cosmology,ratra1988cosmological}.

In this note we have a closer look at the infrared limit of quantum gravity. We pay particular attention to the role of diffeomorphism symmetry. We show that this symmetry enforces cancellations of flow contributions that are crucial for a proper understanding of the infrared limit. At this point one realizes the great advantage of a diffeomorphism invariant flow equation with a single macroscopic metric. In the often used background flow equation \cite{reuter1994effective} it seems rather hard to derive truncations that keep track of these cancellations. The same holds for approaches in flat space that do not employ explicit diffeomorphism symmetry of the effective action \cite{christiansen2014fixed,christiansen2016global,christiansen2015local,denz2016towards}.

The cancellation of contributions from different terms, similar to different loop diagrams in perturbation theory, is crucial for an understanding of the naturalness of a tiny cosmological constant, or of a tiny mass term for the cosmon, the pseudo Goldstone boson of spontaneously broken scale symmetry. In the presence of a fixed point, large individual diagrams do not indicate the ``natural size`` of a quantity. They rather ensure an efficient approach to the fixed point. The ``natural size`` of quantities is determined by their values at the fixed point or in the vicinity of it.

The extreme infrared limit corresponds to vanishing momenta. One therefore evaluates the effective action for metric and scalar fields that do not depend on the position in space and time. In case of diffeomorphism symmetry this focusses on the effective scalar potential in flat space.

In section \ref{Infrared limit of quantum gravity} we show that a suitable truncation of the diffeomorphism invariant flow equation indeed yields a gauge invariant effective action in the infrared limit. In section \ref{Infrared quantum gravity with scalar field} we turn to the infrared limit in the scalar sector and point out the cancellations which occur in the flow of scalar $n$-point functions at zero momentum. We discuss the naturalness of a tiny cosmological constant and cosmon mass. Section \ref{Graviton propagator and vertices at zero momentum} extends this to the graviton propagator and vertices at zero momentum. Our conclusions are discussed in section \ref{Discussion}.

\section{Infrared limit of quantum gravity}
\label{Infrared limit of quantum gravity}

We explore the infrared limit of quantum gravity by solving the flow equation \cite{wetterich1993exact} for the effective average action $\Gamma_{k}$
\begin{equation}\label{A1}
\pt\Gamma_k=\frac12\tr \big\{\pt R_k(\Gamma^{(2)}+R_k)^{-1}\big\},
\end{equation}
with $\pt=k\partial_k$. Here the infrared cutoff $R_k$ suppresses the contribution of modes with small eigenvalues of the covariant Laplacian (or some other suitable differential operator), $-D^2\ll k^2$, while the factor $\partial_t R_k$ removes contributions from 
large eigenvalues $-D^2\gg k^2$ , such that momentum integrals in the one loop expression \eqref{A1} are both infrared and ultraviolet finite. We work within the gauge invariant formalism for a single metric field \cite{wetterich2016gaugeinvariant}. In eq.~\eqref{A1} we have not indicated explicitly the projections on physical modes. We also have omitted the subleading measure contribution. The second functional derivative $\Gamma^{(2)}_k$ depends on the macroscopic metric $g_{\mu\nu}$, which is the argument of the effective average action or flowing action $\Gamma_k$. For $R_k$ a function of suitable covariant derivative operators the r.h.s. of eq. \eqref{A1} is manifully gauge invariant. 

For evaluating the r.h.s. of eq.~\eqref{A1} we employ (in a Euclidean setting) the truncation 
\begin{equation}\label{A2}
\Gamma_k=\int_x\sqrt{g}\left\{V-\frac{M^2}{2}R\right\},
\end{equation}
with $R$ the curvature scalar for the metric $g_\mn$ and $g=\det(g_\mn)$. 
Both $V$ and $M^2$ depend on $k$. They can be functions of scalar fields, and we consider here the situation where both quantities are positive, $V\geq 0$, $M^2\geq 0$. Perturbative non-local corrections typically involve terms $\sim R^2$ or $R_{\mu\nu}R^{\mu\nu}$, with logarithmically running dimensionless coefficients \cite{DONO2}. In the infrared region of $q^2\ll M^2$ (or $-D^2\ll M^2$) these terms play no role for our discussion. For $V>0$ and $V\ll M^4$ they only induce tiny shifts for the location of the onset of instability in momentum space. The shift in the graviton barrier is therefore tiny, and we can omit such terms for our discussion of the gravitational contribution to the flow of $V$. 

\subsection{Gauge invariance of IR-limit}

The infrared limit corresponds to a situation where the physical length scales grow very large. These physical length scales are either given by a curved geometry or by the inverse momentum in some scattering process. In momentum space, the physical length scales act as an effective infrared cutoff for quantum fluctuations, such that fluctuations with momenta $q$ smaller than $q_{\rm phys}$ no longer contribute. The flow with $k$ of vertices and propagators contained in $\Gamma_k$ remains essentially unaffected by the presence of a physical cutoff as long as $k\gg q_{\rm phys}$. On the other hand, this flow effectively stops for $k\lesssim q_{\rm phys}$. For $q_{\rm phys}$ given, for example, by the Hubble parameter of the present universe, we may therefore follow the flow from very large $k$ to $q_{\rm phys}$, neglecting the effects of curved geometry to a good approximation. The infrared limit $q_{\rm phys}\to 0$ translates to the limit $k\to 0$ for the solution of eq. \eqref{A1}. 

The extreme infrared limit (``zero momentum limit'') obtains for an $x$-independent metric $g_{\mu\nu}$, as well as constant scalar fields. This limit is the subject of the present investigation. For a realistic cosmological situation the information about the approach to the extreme infrared limit may be used by stopping the $k$-flow at some characteristic value $k\approx q_{\rm phys}$. The zero momentum limit projects on the first term in eq.~\eqref{A2}. 

We first want to show that the flow equation \eqref{A1} is consistent with the diffeomorphism invariant form for the effective action at zero momentum,
\begin{equation}\label{A3}
\Gamma_V=\int_x\sqrt{g}V,
\end{equation}
verifying explicitly the gauge invariance of the flow equation. For this purpose we have to evaluate the r.h.s. of eq.~\eqref{A1} for arbitrary constant $g_{\mu\nu}$ and to verify that the flow does not induce terms that are not proportional to $\sqrt{g}$.  We can employ $SO(d)$ rotations - the euclidean analogue of the Lorentz transformations - in order to bring any constant $g_{\mu\nu}$ to diagonal form, with eigenvalues $g_\mu>0$,
\begin{equation}\label{A4}
g'_\mn=g_\mu\delta_\mn.
\end{equation}

For an $SO(d)$-invariant cutoff $R_k$ the flow equation is $SO(d)$ invariant provided that $\Gamma$ is $SO(d)$-invariant. It is therefore sufficient to evaluate $\pt\Gamma$ for the metric \eqref{A4}. The dominant contributions to the flow in the infrared arises from the transverse traceless metric fluctuations $t_\mn$ which obey $g^\mn t_\mn=0,~g^{\nu\rho}\partial_\nu t_{\mu\rho}=0$. For positive $V>0$ and in the absence of the IR-cutoff $R_k$ the propagator for these modes shows a tachyonic instability. We will see that the flow is precisely such that this instability is avoided. The dominant IR-contribution of metric fluctuations is associated to a large enhancement near the ``graviton barrier'' \cite{CWGF}, which prevents the singularity to appear. 

The inverse propagator for the transverse traceless metric fluctuations reads in momentum space
\begin{equation}\label{A5}
\Gamma^{(2)\mn\rho\tau}(q)=\sqrt{g}\left(\frac{M^2}{4}q^2-\frac{V}{2}\right)
P^{(t)\mn\rho\tau}(q)\, ,
\end{equation}
with
\begin{equation}\label{A6}
q^2=g^\mn q_\mu q_\nu=\sum_\mu\frac{q^2_\mu}{g_\mu},~
\sqrt{g}=\sqrt{\prod_\mu g_\mu},
\end{equation}
and $P^{\mn\rho\tau}_t$ the projector on transverse traceless metric fluctuations obeying
\begin{equation}\label{A7}
{\rm Tr}P^{(t)}=P^{(t)\mn}_\mn=5.
\end{equation}
In accordance, the infrared cutoff for these fluctuations is chosen as
\begin{equation}\label{A8}
R^{\mn\rt}_k=R_k(q)P^{(t)\mn\rt},~R_k(q)=
\frac{\sqrt{g}M^2k^2}{4}r_k
\left(\frac{q^2}{k^2}\right),
\end{equation}
with $r_k$ a dimensionless function of $q^2/k^2$ to be specified later. In eqs. \eqref{A5}, \eqref{A8} we have omitted the unit operator in momentum space, corresponding to a product of $\delta$-distributions. 

Insertion of eqs. \eqref{A5}, \eqref{A8} into the general flow equation \eqref{A1} yields 
\begin{equation}\label{A9}
\pt(\sqrt{g}V)=5\tilde I_k\left(-\frac{2V}{M^2};g_\mu\right),
\end{equation}
with $\int_q=(2\pi)^{-4}\int d^4 q$ and 
\begin{equation}\label{A10}
\tilde I_k=\frac12\int_q 
\left(q^2+k^2r_k-\frac{2V}{M^2}\right)^{-1}
fk^2r_k,
\end{equation}
where
\begin{equation}\label{A11}
f(x)=2-2\frac{\partial \ln r}{\partial \ln x}, ~x=\frac{q^2}{k^2}.
\end{equation}
The only dependence on $g_\mu$ arises from $q^2$ in eq.~\eqref{A6}. By a rescaling $q_\mu=\sqrt{g_\mu}\tilde q_\mu$ we can achieve
\begin{equation}\label{A12}
q^2=\sum_\mu\tilde q^2_\mu,
\end{equation}
such that the Jacobian,
\begin{equation}\label{A13}
\int\limits_q=\prod_\mu\sqrt{g_\mu}\int_{\tilde q}=\sqrt{g}\int_{\tilde q},
\end{equation}
results in 
\begin{equation}\label{A14}
\tilde I_k=\sqrt{g}I_k.
\end{equation}
The integral
\begin{equation}\label{A15}
I_k=\frac{k^4}{32\pi^2}\int\limits^\infty_0 dx x
\big(p(x)-v)^{-1}f(x)r(x)
\end{equation}
involves
\begin{equation}\label{A16}
p(x)=x+r(x).
\end{equation}
It is independent of $g_\mu$. Besides the factor $k^4$ it only depends on the dimensionless combination $v$, 
\begin{equation}\label{eq:16A} 
v=\frac{2V}{M^2k^2}\, ,
\end{equation}
which may in turn depend on constant scalar fields. Thus $\tilde I_k$ in \eqref{A9} is indeed $\sim \sqrt{g}$. This concludes the argument that the zero-momentum limit of $\Gamma$ is given for all $k$ by the diffeomorphism invariant form \eqref{A3}. 

The simple diffeomorphism invariant form \eqref{A3} of the effective action for constant metric and scalar fields has important consequences. It defines the zero momentum limit of the metric and scalar propagators, as well as all vertices with vanishing external momentum. The zero-momentum limit of the inverse propagator corresponds to the second functional derivative of $\Gamma_V$, while higher derivatives of $\Gamma_V$ yield directly the zero momentum limit of the corresponding one-particle irreducible vertices. 

\subsection{Zero momentum graviton propagator}

In flat space the inverse propagator for transverse traceless metric fluctuations takes by symmetry considerations for all $k$ the general form
\ba\label{B1}
&&G^{-1}_{\mn\rt}(q)=
G^{-1}(q)P^{(t)}_{\mn\rt},\nn\\
&&G^{-1}(q)=-\frac12 V_g
+K_g(q^2),
\ea
with $K_g$ defined such that $K_g(q^2=0)=0$. The zero momentum limit of the propagator $G(q^2=0)$ is then determined by $-2/V_g$. We can compute $G(q)$ by expanding the effective action in second order in the transverse traceless metric fluctuations $t_\mn,g_\mn=\delta_\mn+t_\mn$, 
\begin{equation}\label{B2}
\Gamma_2=\frac12\int_q t^\nu_\mu(-q)G^{-1}(q)t^\mu_\nu(q).
\end{equation}
The zero momentum limit obtains from $\Gamma_V$, employing (for $\bar g^\mn h_\mn=0$)
\ba\label{B3}
&&V\sqrt{\det (\bar g_\mn+h_\mn)}=V
\sqrt{\det \bar g_\mn}
\left(1-\frac14 h^\rho_\tau h^\tau_\rho\right.\\
&&\quad \left. +\frac16 h^\rho_\tau h^\sigma_\rho h^\tau_\sigma 
-\frac18 h^\rho_\tau h^\sigma_\rho h^\eta_\sigma h^\tau_\eta 
+\frac{1}{32}h^\rho_\tau h^\tau_\rho h^\eta_\sigma h^\sigma_\eta +\dots\right)\nn
\ea
Comparison of eqs. \eqref{B1} and \eqref{B2} yields for the quadratic term in eq.~\eqref{B3} the expected result 
\begin{equation}\label{B4}
V_g=V.
\end{equation}
The three-graviton vertex and the four-graviton vertex at zero external momentum are also given by $V$. They correspond to the third and fourth derivative of $\Gamma_V$, as given by the terms cubic and quartic in $h^\nu_\mu$ in eq.~\eqref{B3}. 

\subsection{Flow of effective scalar potential}

The contribution of the transverse traceless metric fluctuations to the flow of $V$ for arbitrary values of constant scalar fields can be evaluated for $g_\mn=\delta_\mn$ in eq.~\eqref{A9}, resulting in the flow equation for $V$ \cite{CWGF}
\begin{equation}\label{A16A}
\pt V=\frac{5k^4}{32\pi^2}\int dx xf(x)r(x)\big(p(x)-v\big)^{-1}.
\end{equation}
One obtains for the dimensionless combination $v$
\begin{equation}\label{A17}
\pt v=-2v
-\frac{\partial_tM^2}{M^2}v
+\frac{5k^2}{8\pi^2 M^2}l_0(-v),
\end{equation}
with ``threshold function''
\begin{equation}\label{A18}
l_0(w)=\frac12\int\limits^\infty_0 dx x\big(p(x)+w\big)^{-1}f(x)r(x).
\end{equation}
The threshold functions depend on the choice of the cutoff function $r_k(x)$ and have been discussed extensively in the literature \cite{LiCu,BTW,CWSP,litim2001optimized}, with $l_n(w)=l^4_n(w)$. 

We may consider a choice of IR-cutoffs such that $p(x)$ has a minimum at $\bar x$, with expansion close to the minimum
\begin{equation}\label{A19}
p(x)=\bar p+a(x-\bar x)^2.
\end{equation}
Close to the graviton barrier only a narrow range in $|x-\bar x|$ makes a substantial contribution to the integral \eqref{A18}. We approximate $x\approx \bar x$ and 
\begin{equation}\label{A20}
f(x)r(x)\approx f(\bar x)r(\bar x)=2\bar s.
\end{equation}
This yields for $v$ in the vicinity of $\bar p$
\begin{equation}\label{A21}
l_0(-v)=\frac{\pi\bar x\bar s}{{\sqrt a}}(\bar p-v)^{-1/2},
\end{equation}
and we conclude that the threshold function diverges for $v\to \bar p$. The flow of $v$ prevents $v$ to exceed $\bar p$ - this is the ``graviton barrier''. 

We do not attempt to compute the flow of $M^2$ in this note. As an example we consider a scenario where $|\partial_t\ln M^2|\ll 1$ such that $M^2$ becomes effectively a $k$-independent quantity. This may not be the correct infrared behavior of $M^2$, but it illustrates well our purposes. For constant $M^2$ one finds for $v$ close to $\bar p$
\begin{equation}\label{A22}
\pt v=-2\bar p+\frac{\bar e k^2}{M^2}(\bar p-v)^{-1/2},
\end{equation}
where 
\begin{equation}\label{A23}
\bar e=\frac{5\bar x\bar s}{8\pi\sqrt{a}}.
\end{equation}
The flow of $v$ is attracted for decreasing $k$ to an approximate partial IR-fixed point for which the r.h.s. of eq. \eqref{A22} vanishes,
\begin{equation}\label{A24}
v=\bar p-
\left(\frac{\bar ek^2}{2\bar p M^2}\right)^2,~V=\frac{\bar p k^2M^2}{2}-
\frac{\bar e^2k^6}{8\bar p^2M^2}.
\end{equation}
For small values of $k^2/M^2$ the ratio $V/M^2$ comes very close to a universal IR-value
\begin{equation}\label{A25}
\frac{V}{M^2}\approx \frac{\bar p k^2}{2}.
\end{equation}
It vanishes for $k\to 0$.

\section{Infrared quantum gravity with scalar field}
\label{Infrared quantum gravity with scalar field}

In this section we investigate in more detail the flow of derivatives of the effective potential. This will reveal important cancellations of individual contributions, due to gauge symmetry and the presence of a fixed point. 

\subsection{Flow of $\chi$-dependence of scalar potential}

Let us next assume that $V$ depends on a single scalar field $\chi$, without specifying its normalization. We want to understand the flow of the $\chi-$dependence of $V(\chi)$, and concentrate again on the dominant contribution from the transverse traceless metric fluctuations. 
The flow of $\partial V/\partial \chi$ obtains by taking a $\chi$-derivative of the flow equation for $V$,
\begin{equation}\label{B5}
\pt V=\frac52 \int_q \pt R_k(q)
P^{-1}_k(q),
\end{equation}
with 
\begin{equation}\label{31a}
P_k(q)=\frac{M^2q^2}{4}-\frac V2+R_k(q).
\end{equation}
The $\chi$-derivative of eq.~\eqref{B5} yields
\ba\label{B6}
\pt\frac{\partial V}{\partial \chi}=-\frac52\int_q\pt R_k
P^{-2}_k(q)
\left(\frac{q^2}{4}\frac{\partial M^2}{\partial\chi}-\frac12\frac{\partial V}{\partial \chi}\right)+\Delta^{(1)}_R,\nn\\
\ea
with 
\ba\label{B7}
\Delta^{(1)}_R= \frac52\int_q\left[\pt
\left(\frac{\partial R_k}{\partial\chi}\right)
P^{-1}_k(q)-
\pt R_k
P^{-2}_k(q)
\frac{\partial R_k}{\partial\chi}\right].\nn\\
\ea

\begin{figure}[h!tb]
\includegraphics[width=0.2\textwidth]{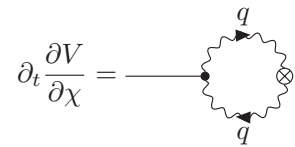}
\caption{Graphical representation of a contribution to the flow of $\partial V/\partial \chi$. For the other contributions the scalar line ends in the $R_{k}$-insertion.}
\label{Iqg_Fig1}
\end{figure}

The first term in eq.~\eqref{B6} has a standard representation as a one-loop diagram, shown in Fig. \ref{Iqg_Fig1}.


\noindent
Here the curled line denotes a graviton propagator, the cross the insertion of $\pt R_k$ and the dot the scalar-graviton-graviton cubic vertex. According to the truncation \eqref{A2} the vertex for a zero-momentum scalar and two gravitons with momenta $q$ and $-q$ is given as
\begin{equation}\label{B8}
V_3(0,q,-q)=\frac{q^2}{4}\frac{\partial M^2}{\partial \chi}-\frac12\frac{\partial V}{\partial\chi}.
\end{equation}
Insertion of these expressions in the graph of Fig. \ref{Iqg_Fig1}, together with the statistical factor $5/2$ accounting for the five degrees of freedom, yields indeed the first term in eq. \eqref{B6}.

The second term $\Delta^{(1)}_R$ arises from the dependence of $R_k$ on $\chi$. It reflects our implicit definition of the effective average action where $R_k$ depends on the macroscopic fields $\chi$ and $g_\mn$, rather than on some fixed background fields. In particular, for the choice \eqref{A8} one has $\partial_{\chi} R_k=\partial_{\chi}(\ln M^2)R_k$. The additional term ensures that derivatives commute,
\begin{equation}\label{B9}
\pt\frac{\partial V}{\partial \chi}=\frac{\partial}{\partial\chi}\pt V.
\end{equation}
This is a crucial feature of a closed gauge invariant flow equation for a single metric and scalar field \cite{wetterich2016gaugeinvariant}. It distinguishes the present approach from the background field formalism. 

\subsection{Infrared enhancement for $\chi$-derivatives}

For $M^2$ independent of $k$ one obtains for the $\chi$-derivative of the dimensionless ratio $v$
\begin{equation}\label{B10}
\pt \left(\frac{\partial v}{\partial\chi}\right)=(A_{v1}-2)
\frac{\partial v}{\partial \chi}-
\frac{5k^2}{8\pi^2M^2}l_0(-v)
\frac{\partial\ln M^2}{\partial \chi},
\end{equation}
with 
\begin{equation}\label{B11}
A_{v_1}=\frac{5k^2}{8\pi^2M^2}l_1(-v),
\end{equation}
where the appropriate threshold function reads
\begin{equation}\label{B12}
l_1(w)=\frac12\int dxxf(x)r(x)\big(p(x)+w\big)^{-2}=-\frac{\partial}{\partial w}l_0(w).
\end{equation}
We are interested here in the range of $v$ for which the infrared enhancement of the graviton fluctuations occurs, e.g. for $v$ near $\bar p$.

Typically, the flow equations for derivatives of $\Gamma$, e.g. inverse propagators and vertices, contain a higher number of graviton propagators in the loop. This is reflected by a stronger divergence of individual diagrams in the vicinity of the graviton barrier. In our case this is manifest by the stronger divergence of $l_1(-v)$ as compared to $l_0(-v)$,
\begin{equation}\label{B13}
l_1(-v)=\frac{\pi\bar x\bar s}{2\sqrt{a}}(\bar p-v)^{-3/2}=\frac{l_0(-v)}{2(\bar p-v)}.
\end{equation}
For the approximate scaling solution \eqref{A24} this results in a huge enhancement of $l_1/l_0$ for small $k^2/M^2$,
\begin{equation}\label{B14}
\frac{l_1(-v)}{l_0(-v)}=\frac12
\left(\frac{2\bar p M^2}{\bar e k^2}\right)^2.
\end{equation}
One may wonder what are the effects of this additional huge infrared enhancement, in particular if it could invalidate our discussion of the approach to the scaling form of $V$. 

According to eq.~\eqref{B10} the flow of $\partial v/\partial \chi$ is attracted for $k\to 0$ towards the approximate partial fixed point
\begin{equation}\label{B15}
\frac{\partial v}{\partial\chi}=\frac{l_0(-v)}{l_1(-v)}
\frac{\partial \ln M^2}{\partial\chi}.
\end{equation}
For a tiny ratio $l_0/l_1$ the dimensionless combination $v$ becomes therefore almost independent of $\chi$. The additional IR-enhancement leads to a rapid flow of $v(\chi)$ towards a $\chi$-independent value. As it should be, this corresponds to the scaling solution \eqref{A24}, which implies directly
\begin{equation}\label{B16}
\frac{\partial v}{\partial \chi}=2
\left(\frac{\bar e k^2}{2\bar pM^2}\right)^2
\frac{\partial \ln M^2}{\partial \chi}.
\end{equation}

We observe that the ``IR-enhancement'' of diagrams with a higher number of graviton propagators in the loop affects only the small difference between $V$ and the scaling form
\begin{equation}\label{B17}
V_c(\chi)=\frac{\bar pk^2M^2(\chi)}{2},
\end{equation}
which corresponds to the $\chi$-independent value $v_c=\bar p$. For 
\begin{equation}\label{B18}
M^2(\chi)=c\chi^\alpha k^{2-\alpha}
\end{equation}
the derivative $\partial V_c/\partial\chi$ shows no trace of an additional IR-enhancement. It follows the order of magnitude estimate $\partial V/\partial\chi\approx V/\chi$, e.g.
\begin{equation}\label{B19}
\chi\frac{\partial V_c}{\partial \chi}=\alpha V_c.
\end{equation}

\subsection{Flow of scalar propagator and IR-cancellation}

\begin{figure}[h!tb]
\includegraphics[width=0.45\textwidth]{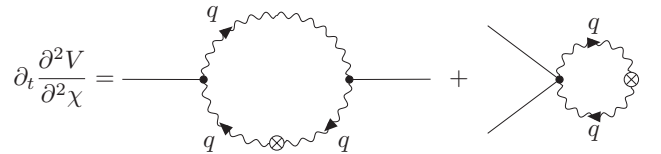}
\caption{Graphical representation of contributions to the flow of $\partial^2 V/\partial \chi^2$. For other contributions scalar lines end in $R_{k}$-insertions.}
\label{Iqg_Fig2}
\end{figure}

We next address the flow of the scalar propagator at zero momentum. We discuss the contributions of various diagrams separately. This will reveal strong cancellation effects between different diagrams. The flow of the inverse scalar propagator at zero momentum (mass term), $\partial^2V/\partial\chi^2$, obeys
\begin{equation}\label{B20}
\pt \frac{\partial^2 V}{\partial \chi^2}=\zeta^{(3)}_m+\zeta^{(4)}_m+\Delta^{(2)}_R,
\end{equation}
with 
\ba\label{21}
\zeta^{(3)}_m=5\int_q\pt R_k
P^{-3}_k(q)
\left(\frac{q^2}{4}\frac{\partial M^2}{\partial\chi}-\frac12\frac{\partial V}{\partial \chi}\right)^2\, ,
\ea
and 
\begin{equation}\label{B22}
\zeta^{(4)}_m=-\frac52\int_q\pt R_k
P^{-2}_k(q)\left(\frac{q^2}{4}\frac{\partial^2 M^2}{\partial \chi^2}-\frac12\frac{\partial^2 V}{\partial \chi^2}\right)\, ,
\end{equation}
while $\Delta^{(2)}_R$ denotes terms involving $\chi$-derivatives of $R_k$. The graphical representation of the first two terms is shown in Fig. \ref{Iqg_Fig2}.


Evaluating the contribution of the diagram with the squared cubic vertex yields for the region close to the graviton barrier
\begin{equation}\label{B23}
\zeta^{(3)}_m=\frac{5}{2\pi^2 M^4}
\left(\frac{k^2\bar x}{4}
\frac{\partial M^2}{\partial\chi}-\frac12
\frac{\partial V}{\partial \chi}\right)^2 l_2(-v),
\end{equation}
with threshold function
\ba\label{B24}
l_2(w)&=&\int dxx f(x) r(x)\big(p(x)+w\big)^{-3}\nn\\
&=&-\frac{\partial l_1}{\partial w}=
\frac{3\pi\bar x\bar s}{4\sqrt{a}}(\bar p+w)^{-5/2}.
\ea
For $V$ near $V_c$ one obtains
\begin{equation}\label{B25}
\zeta^{(3)}_m =\frac{3\bar e(\bar x-\bar p)^2k^4}{16}(\bar p-v)^{-\frac52}
\left(\frac{\partial\ln M^2}{\partial\chi}\right)^2.
\end{equation}
Inserting for $\bar p-v$ the scaling solution \eqref{A24} produces for large $M/k$ a huge enhancement factor
\begin{equation}\label{B26}
(\bar p-v)^{-\frac52}=\left(\frac{2\bar pM^2}{\bar ek^2}\right)^5.
\end{equation}

A naive estimate of the order of $\partial V/\partial \chi^2$ by the contribution of the first diagram in Fig. \ref{Iqg_Fig2} is wrong by many orders of magnitude. For $M^2$ of the form \eqref{B18} one has $(\partial\ln M^2/\partial\chi)^2=\alpha^2/\chi^2$. With this form of $M^2$ the correct value of $\partial^2 V/\partial\chi^2$ is given by
\begin{equation}\label{B27}
\frac{\partial^2 V_c}{\partial\chi^2}=
\frac{\alpha(\alpha-1)\bar p k^2 M^2}{2\chi^2}. 
\end{equation}
The ratio
\begin{equation}\label{B28}
\zeta^{(3)}_m\left/\frac{\partial^2 V_c}{\partial\chi^2}=
\frac{3\alpha(\bar x-\bar p)^2}{4(\alpha-1)}
\left(\frac{2\bar pM^2}{\bar e k^2}\right)^4\right.
\end{equation}
is huge for large $M^2/k^2$. The reason for the failure of the naive order of magnitude estimate is the neglection of two important properties. The first is that the flow of $\partial V^2/\partial\chi^2$ arises from the second derivative of the flow of $V$, which involves a less IR-enhanced diagram. This entails partial cancellations between different diagrams, e.g. between $\zeta^{(3)}_m,\zeta^{(4)}_m$ and $\Delta^{(R)}_2$. The second is the strong attraction to the approximate IR-fixed point, which also enforces cancellations.

The sum of all contributions to the flow of $\partial V/\partial\chi^2$ can be inferred by taking a $\chi$-derivative of eq.~\eqref{B10},
\ba\label{B29}
&&\pt\left(\frac{\partial^2 v}{\partial\chi^2}\right)=
\frac{5k^2}{8\pi^2M^2}
\left\{ l_2(-v)
\left(\frac{\partial v}{\partial\chi}\right)^2\right.\\
&&\quad+l_1(-v)\left[\frac{\partial^2 v}{\partial\chi^2}-2\frac{\partial\ln M^2}{\partial\chi}\frac{\partial v}{\partial\chi}\right]\nn\\
&&\quad \left. +l_0(-v)
\left[\left(\frac{\partial\ln M^2}{\partial\chi}\right)^2-
\frac{\partial^2\ln M^2}{\partial\chi^2}\right]\right\}
-2\frac{\partial^2v}{\partial\chi^2}\nn.
\ea
We observe again the strong IR-enhancement factor $l_2(-v)$ from the diagrams with three graviton propagators. Close to the scaling solution this is multiplied by the tiny factor $(\partial v/\partial\chi)^2$ according to eq.~\eqref{B16}. The combination
\begin{equation}\label{B30}
\frac{5k^2}{8\pi^2M^2}l_2(-v)
\left(\frac{\partial v}{\partial\chi}\right)^2=6\bar p
\left(\frac{\partial\ln M^2}{\partial\chi}\right)^2
\end{equation}
contains no longer any strong IR-enhancement, in contrast to the combination $\zeta^{(3)}_m$ in eqs. \eqref{B25}, \eqref{B26}, which only includes a particular diagram. The contribution to the flow \eqref{B30} is of the same order as other terms. Inserting the scaling solution for $v$ and $\partial v/\partial\chi$, eq.~\eqref{B29} becomes
\ba\label{B31}
\pt \left(\frac{\partial^2v}{\partial\chi^2}\right)&=&
\left[\bar p\left(\frac{2\bar p M^2}{\bar ek^2}\right)^2-2\right]
\frac{\partial^2 v}{\partial\chi^2}\nn\\
&&-2\bar p\left[\frac{\partial^2\ln M^2}{\partial\chi^2}-2
\left(\frac{\partial\ln M^2}{\partial\chi}\right)^2\right].
\ea
For large $M^2/k^2$ the second derivative of $v$ is attracted very rapidly towards the approximate partial fixed point
\begin{equation}\label{B32}
\frac{\partial^2 v}{\partial\chi^2}=2
\left(\frac{\bar ek^2}{2\bar pM^2}\right)^2
\left[\frac{\partial^2\ln M^2}{\partial\chi^2}-2
\left(\frac{\partial\ln M^2}{\partial\chi}\right)^2\right].
\end{equation}
As it should be, the latter coincides with the $\chi$-derivative of eq.~\eqref{B16}.

\subsection{Cancellations and naturalness}

We have made this somewhat lengthy explicit demonstration in order to show that a naive estimate of the ``natural order of magnitude'' of a quantity by the evaluation of a typical contributing loop diagram can be wrong by a huge factor. Infrared fixed points can lead to almost complete cancellations of terms, inducing small quantities whose symmetry origin is not immediately visible. Part of the properties can be understood by the realization of dilatation or scale symmetry at the fixed point \cite{CWEP}. Another part is related to the particular approach towards the fixed point, e.g. the corresponding ``anomalous dimensions'' characterizing the stability matrix. 

These properties are relevant for two important issues in particle physics and cosmology, namely the naturalness of a small cosmological constant and a small mass for the cosmon - the scalar field responsible for dynamical dark energy. For both cases naive order of magnitude estimates suggest that tiny values are unnatural. We will see that the fixed point behavior precisely leads to such small values. Our findings confirm the general discussion of the consequences of an IR-fixed point for the issue of naturalness in ref.~\cite{CWEP}. For a naive estimate large individual contributions to the flow seem to imply large ``natural'' values of the corresponding flowing coupling. In the presence of an IR-fixed point large contributions to the flow induce a particular rapid approach to the fixed point and therefore particular efficient cancellations of individual terms. This is the reason why the scaling form of the potential $V_c$ is realized with a very high accuracy for small $k^2/M^2$.

\section{Graviton propagator and vertices at zero momentum}
\label{Graviton propagator and vertices at zero momentum}

For the graviton propagator and vertices at zero momentum similar considerations apply. In addition, an important particular feature is related to the presence of diffeomorphism symmetry. We have already established that the inverse graviton propagator at zero momentum $V_g$ equals the effective potential $V$, as required by diffeomorphism symmetry if a derivative expansion for the effective action is valid for the lowest order term. It is interesting to investigate again various individual contributions to the flow equation for $V_g$ and to see their partial cancellations.

For this purpose we evaluate the flow of the $\Gamma$ for a constant diagonal traceless metric
\begin{equation}\label{C1}
g_\mn={\rm diag}(1,1+g_1,1-g_1,1),
\end{equation}
e.g.
\begin{equation}\label{C2}
\Gamma=\int_x{\cal L}_0(g_1).
\end{equation}
The inverse graviton propagator at zero momentum is given by the second derivative of ${\cal L}_0$,
\begin{equation}\label{C3}
\frac{\partial^2{\cal L}_0}{\partial g^2_1}|_{g_1=0}=-V_g.
\end{equation}
For the truncation \eqref{A2} the flow equation for ${\cal L}_0(g_1)$ reads explicitly
\ba\label{C4}
\pt \cl_0&=&\frac52\int_q\pt R_k(q)
\left[\frac{M^2\sqrt{1-g^2_1}}{4}\right.
\left(q^2_0+\frac{q^2_1}{1+g_1}\right.\nn\\
&&\left.+\frac{q^2_2}{1-g_1}+q^2_3\right)-\frac{V\sqrt{1-g^2_1}}{2}+R_k(q)\Bigg]^{-1}
\ea

\begin{figure}[h!tb]
\includegraphics[width=0.45\textwidth]{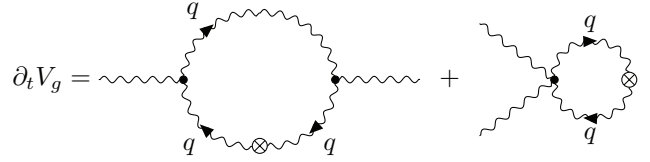}
\caption{Contributions to the flow of the inverse graviton propagator. For other contributions graviton lines end in $R_{k}$-insertions.}
\label{Iqg_Fig3}
\end{figure}

Taking two derivatives with respect to $g_1$ one finds
\begin{equation}\label{C5}
\pt \frac{\partial^2\cl_0}{\partial g^2_1}|_{g_1=0}=\zeta^{(3)}_g+\zeta^{(4)}_g+\Delta^{(g2)}_R,
\end{equation}
with 
\ba\label{C6}
\zeta^{(3)}_g=5\int_q\pt R_k(q)P^{-3}_k(q)
\left[\frac{M^2}{4}(q^2_2-q^2_1)\right]^2,
\ea
and 
\ba\label{C7}
&&\zeta^{(4)}_g=-\frac52\int_q\pt R_k(q)P^{-2}_k(q)\nn\\
&&\qquad \times\left[\frac{M^2}{4}(q^2_1+q^2_2
-q^2_0-q^2_3)+\frac V2\right],
\ea
with $q^2=\sum_\mu q^2_\mu$. Again, the part $\Delta^{(g2)}_R$ involves derivatives of the cutoff functions $R_k$ with respect to $g_1$. The contributions $\zeta^{(3)}_g$ and $\zeta^{(4)}_g$ are depicted as graphs in Fig. \ref{Iqg_Fig3}.


We observe that the part $\zeta^{(3)}_g$ with two three-graviton vertices does not have a part involving $V$ in the last bracket in eq.~\eqref{C6}. This holds despite the presence of a term $\sim h^3$ in eq.~\eqref{B3}. For the part $\zeta^{(4)}_g$ proportional to the four-graviton vertex the momentum integration retains only the term $\sim V/2$ in the last bracket of eq.~\eqref{C7}, resulting in 
\begin{equation}\label{C8}
\zeta_g^{(4)}=-\frac{5Vk^2}{16\pi^2M^2}l_1(-v).
\end{equation}
For the scaling solution close to the graviton barrier one finds
\begin{equation}\label{C9}
\zeta^{(4)}_g=-\frac{\bar p V}{2}
\left(\frac{2\bar p M^2}{\bar ek^2}\right)^2
\end{equation}
A naive estimate of the order of magnitude of $V_g$ by the size of the flow contribution $\zeta^{(4)}_g$ would result in an overestimate by a huge factor $\sim(M^2/k^2)^2$. This is even worse for $\zeta^{(3)}_g$ which involves a factor $(M^2/k^2)^4$. 

For the complete calculation the term $\Delta^{(g2)}_R$ in eq.~\eqref{C5} is important. For our realization of diffeomorphism symmetry the cutoff function involves the macroscopic metric $g_\mn$ in the form of covariant derivatives and by the multiplicative factor $\sqrt{g}$. As a result one finds by a suitable momentum rescaling (as discussed above) the simple relation
\begin{equation}\label{C10}
\pt\cl_0(g_1)=\sqrt{(1-g^2_1)}\pt V,
\end{equation}
with $\pt V$ given by eq.~\eqref{A16A} being independent of $g_1$. This implies directly the expected result
\begin{equation}\label{C11}
\pt V_g=-\pt\frac{\partial^2\cl_0}{\partial g^2_1}|_{g_1=0}=\pt V.
\end{equation}
The presence of the term $\Delta_{R}^{(g2)}$ is crucial for this argument. Omitting it, the rescaling of momenta no longer permits the tremendous simplification. In other words, if we omit the dependence of $R_{k}$ on the macroscopic metric through covariant derivatives, strong violations of diffeomorphism symmetry would occur. 

Also the flow of a graviton vertices at zero momentum is given by $\pt V$, according to the relation for homogeneous $g_\mn$
\begin{equation}\label{C12}
\pt\cl_0(g_\mn)=\sqrt{\det (g_\mn)}\pt V.
\end{equation}

We conclude that a proper implementation of diffeomorphism symmetry - in our case by the dependence of the cutoff $R_k$ on the macroscopic fields $g_\mn$ and $\chi$ - is crucial for a reliable result in the vicinity of the graviton barrier. Otherwise spurious large infrared enhancement factors can lead easily to huge errors. 

\section{Discussion}
\label{Discussion}

 Let us discuss our findings in the context of scalar tensor theories. For such theories of a scalar $\chi$ coupled to the metric, the functional flow equations based on the background field method or expansions around flat space have been discussed in refs.~\cite{percacci2003asymptotic,narain2010renormalization,henz2013dilaton,henz2017scaling}. Away from the strong non-perturbative infrared flow our approach based on the diffeomorphism invariant flow equation is expected to yield similar results. In the vicinity of the singularity, however, the strong cancellation effects discussed in the present note require a detailed understanding of the consequences of gauge symmetry, for which our method is particularly well suited.
 
Our central finding concerns the graviton barrier for the scalar effective potential $V(\chi)$, cf. eq.~\eqref{A25},
\begin{equation}\label{ZA} 
V(\chi)\leq  \frac{\bar{p} k^2 M^{2}(\chi)}{2},
\end{equation}
together with the diffeomorphism invariant form \eqref{A3} of the zero momentum limit. This implies that the dimensionless ratio 
\begin{equation}\label{73A}
\lambda=\frac{V}{M^4}\leq \frac{\bar p k^2}{2M^2}
\end{equation}
vanishes for $k\to 0$ if $\lambda\geq 0$. The ratio $\lambda$ plays the role of a dimensionless frame invariant cosmological constant \cite{wetterich1988cosmology}. In the Einstein frame the cosmological constant becomes $\lambda_E=\lambda M^4$ and therefore vanishes $\sim k^2$ or faster for $k\to 0$. We have only derived an upper bound on $\lambda$. A behavior $\lambda_E\sim k^4$ \cite{henz2013dilaton,henz2017scaling} is perfectly compatible with this bound. 

The flow of $v$ typically has an UV- and an IR-partial fixed point. Indeed, a simple structure for the zeros of $\beta_v=\partial_t v$ follows if $\beta_v$ is negative for some region of $v$ due to the term $-2v$. For $v\to \bar p$ the function $\beta_v$ always becomes positive due to the enhanced graviton contribution near the barrier. On the other hand, for $v\to -\infty$ the term $-2v$ dominates such that $\beta_v$ is again positive for large negative $v$. The necessary zeros of $\beta_v$ inbetween the limits correspond to the partial fixed points. If we consider any fixed $\bar k$ and start for large $k$ close enough to the UV-fixed point, the infrared instability is not reached at $k=\bar k$. The bound \eqref{ZA} is obeyed nevertheless, such that for $V>0$ the potential cannot increase faster than $M^2$ as $\chi$ increases. 

A general trajectory as a function of $k$ is a crossover from the UV-fixed point to the IR-fixed point. One may associate $k$ qualitatively with a physical IR-cutoff $q_{\rm phys}$ by external momenta in some scattering process or from a curved background geometry. More precisely, if $k$ gets smaller than $q_{\rm phys}$ the flow with $k$ stops, such that in the limit $k\to 0$ one replaces $v(\chi/k)$ by $v(\chi/q_{\rm phys})$. If no intrinsic scale is present, the crossover of $v$ as a function of $\chi$ corresponds to the scaling solution for a scale invariant quantum effective action. The whole scaling solution can be associated to the UV-fixed point. 

From the  point of view of the UV-fixed point the ratio $v$ is a relevant parameter. If this parameter flows away from the UV-fixed point this will induce some characteristic intrinsic scale $\bar \mu$ that indicates that $v$ is no longer close to its fixed point behavior. As long as $k>\bar\mu$ the flow is close to the fixed point and one can neglect the presence of $\bar\mu$. Let us make the assumption (realized in many crossover situations) that the flow of $v$ with $k$ stops for $k<\bar\mu$. For $k\to 0$ we can then replace $v(\chi/k)$ by $v(\chi/\bar \mu)$, such that 
\begin{equation}\label{73B}
V=\frac12\bar\mu^2M^2 v(\chi/\bar\mu).
\end{equation}
This interpolates between
\begin{equation}\label{73C}
V_{\rm IR}=\frac{\bar p}{2}\bar\mu^2M^2
\end{equation}
for $v$ near $\bar p$, and 
\begin{equation}\label{73D}
V_{\rm UV}\sim \bar \mu^4M^2/\chi^2
\end{equation}
for $v$ in the region where $\beta_v\approx -2v$. For $\chi$ in the region where eq. \eqref{73D} applies the IR-enhancement of the graviton fluctuations is not effective. A qualitative form of the effective potential is given by 
\begin{equation}\label{75E}
V(\chi)=\mu^2M^2(\chi)+c\mu^4M^2(\chi)/\chi^2,
\end{equation}
where $\mu^2=\bar p \bar \mu^2/2$ and the coefficient $c$ depends on the details how the flow with $k$ is stopped for $k<\bar \mu$. 

If $M^{2}$ increases monotonically with increasing $\chi$, one may choose a normalization of $\chi$ where $M^{2}=\chi^{2}$. The effective action for the scalar-graviton system becomes then
\begin{eqnarray}\label{ZB} 
\Gamma&=&\int_{x}\, \sqrt{g}\left \{-\dfrac{\chi^{2}}{2}R+\mu^{2}\chi^{2}+c\mu^4\right.\nonumber\\
&&\left.+\dfrac{1}{2}\bigl (B(\chi/\mu)-6\bigl )\p^{\mu}\chi\p_{\mu}\chi\right \}.
\end{eqnarray}
Here, the form of the scalar kinetic term, encoded in the dimensionless function $B(\chi/\mu)$, has not yet been computed so far. Stability requires $B(\chi/\mu)\geq 0$. With a specification of the qualitative behavior of $B$, eq.~\eqref{ZA} constitutes the effective action of variable gravity \cite{wetterich2014variable}. Realistic cosmology, with $\chi$ playing the role of the inflaton in early cosmology, and the cosmon of dynamical dark energy in late cosmology, can be obtained \cite{wetterich2015inflation,rubio2017emergent} for a suitable flow of the dimensionless kinetic function $B(\chi/\mu)$ from a large UV-value for small $\chi/\mu$ to a small IR-value for large $\chi/\mu$. For the cosmological solutions of the field equations derived from eq. \eqref{ZB} one finds that $R$ is of the order $\mu^2$ for all times. For this type of solutions the graviton instability is cured on-shell by the geometry. A nonzero value of $V$, as implemented by the stop of the flow for $k<\bar \mu$, does not lead to unstable behavior of the cosmological solution. 

The generic solution of the cosmological constant problem for an effective action of the type \eqref{ZB} is very simple. It suffices that cosmology is described by a ``runaway solution'' where $\chi\to\infty$ for $t\to\infty$. The cosmological constant in the Einstein frame is given (for $M$ the fixed reduced Planck mass) by 
\begin{equation}\label{73C-1}
\lambda_E=\frac{V}{\chi^4}\cdot M^4=
\frac{\mu^2M^4}{\chi^2}+
\frac{c\mu^4M^4}{\chi^4}.
\end{equation}
It vanishes in the infinite future since $\chi\to\infty$. At present $\lambda_E$ still differs from zero since $\chi=M$, explaining a non-vanishing dynamical dark energy. No small dimensionless parameter is needed since $\mu$ is the only intrinsic scale. The present dark energy density is tiny since the universe is very old and the ratio $\chi/\mu$ therefore very large. This basic mechanism was underlying the prediction \cite{wetterich1988cosmology} of a homogeneously distributed dynamical dark energy of the same order of magnitude as dark matter, long before its discovery. 

The derivation of the graviton barrier in this note has made two important assumptions. The first states that the inverse graviton propagator for momenta $q^{2}$ in the vicinity of $k^{2}$ is reasonably well approximated by eq.~\eqref{A5}. In the presence of the strong IR-flow close to the singularity the validity of a derivative expansion in the gravitational sector is not guaranteed. The effective gravitational action may contain non-local terms. The definition of the parameter $M^{2}$ may become ambiguous. For our purposes it should be defined by the graviton propagator at non-zero momentum, inserting $q^{2}=\bar{p}k^{2}$ in eq.\eqref{A5}. The next important step concerns therefore the understanding of the graviton propagator in the range $q^{2}\approx k^{2}$, in addition to the behavior at $q^{2}=0$ discussed in the present paper.

The second assumption has neglected the flow of $M^{2}$ with $k$. The additional term $\sim \p_{t}\ln(M^{2})$ in eq.~\eqref{A17} cannot remove the singular behavior of the last term as the graviton barrier is approached for $v\raw\bar{p}$. In principle, it is not yet excluded that $\p_{t}\ln(M^{2})$ also becomes singular for $v\raw\bar{p}$, or even for $v<\bar{p}$. Even in this case, and even for a more complicated non-local form of the gravitational effective action in the IR-domain, we expect that a proper treatment of the flow never enters a parameter range where the effective action becomes unstable. This will always imply an upper bound on $v$ and therefore some form of a graviton barrier.

In this note we have not discussed in detail how cosmology provides for a dynamical solution of the cosmological constant problem. This has been done in numerous papers on variable gravity in the past. The reader may find detailed investigations of observable consequences as early dark energy in these papers \cite{wetterich2015inflation,rubio2017emergent,wetterich1988cosmology,wetterich2014variable}. The emphasis of the present paper is more on technical aspects, concerning the issues of cancellations and naturalness \cite{CWEP}. We find important cancellations between individual contributions to the cosmological constant or effective scalar potential. They are due to symmetry and the presence of an IR-fixed point. Rather than being in contradiction to the naturalness of a tiny value of the cosmological constant or present dark energy density, these cancellations are a direct consequence of the general structure of the flow near fixed points. 

We believe that our work contains general lessons for the discussion of naturalness of small parameters. We have demonstrated that in the vicinity of a fixed point the naive estimate of the order of magnitude of some quantity by the size of an individual loop contribution can be erroneous by many orders of magnitude. This naive estimate is, however, the approach often taken in discussions of the naturalness of a small cosmological constant. The present work demonstrates that the proper understanding of the flow near fixed points is crucial for estimates of the ``natural`` range of values for the cosmological constant.

\medskip\noindent
{\em Acknowledgment}

\noindent
The author would like to thank J.~Pawlowski for comments and discussion. This work is supported by ERC-advanced grant 290623 and the DFG Collaborative Research Centres ``TRR33 (Dark Universe)'' and ``SFB  1225  (ISOQUANT)''.


\bibliography{Infrared_limit_of_quantum_gravity}

\end{document}